\documentclass{article}

\usepackage{arxiv}
\usepackage{subfigure}
\usepackage{amsmath,amssymb,euscript,yfonts,psfrag,latexsym,dsfont,graphicx}
\usepackage{bbm,color,amstext,wasysym,parskip,balance,cite}
\usepackage{comment}

\graphicspath{{./},{./figures/}}

\newcommand{\range}{\operatorname{range}}
\newcommand{\thespan}{\operatorname{span}}
\usepackage[ruled,vlined]{algorithm2e}
\newtheorem{prop}{Proposition}

\title{The Challenge of Small Data:\\ Dynamic Mode Decomposition, Redux
}

\author{
  Amirhossein Karimi \\
  Department of Mechanical\\
  and Aerospace Engineering\\
  University of California, Irvine\\
  California, USA \\
  \texttt{amirhosk@uci.edu} \\
   \And
 Tryphon T.\ Georgiou \\
  Department of Mechanical\\
  and Aerospace Engineering\\
  University of California, Irvine\\
  California, USA \\
  \texttt{tryphon@uci.edu} \\
}

\begin{document}
\maketitle

\begin{abstract}
We revisit the setting and the assumptions that underlie the methodology of Dynamic Mode Decomposition (DMD) in order to highlight caveats as well as potential measures of when the applicability is warranted.
\end{abstract}


\section{Introduction}
Whereas the topic of ``big data'' dominates current headlines in research publications and popular news analyses alike,
the perennial challenge of obtaining reliable models with only limited observation records persists in a wide range of time series applications.
Indeed, one often hears the admission from practitioners that the problem is not ``big data'' but ``small data.'' A case in point is that of time series of flow fields where a exceedingly high-dimension state is observed, or partially observed, albeit over a relatively short time window.  It is precisely for these types of applications that Dynamic Mode Decomposition (DMD) and related frameworks were conceived to address \cite{kutz2016dynamic,williams2015data}.

DMD, as introduced by Schmidt \cite{schmid2010dynamic}, is a formalism to identify dominant modes in a high-dimensional time series $x_t\in \mathbb R^N$, $t\in\{1,2,\ldots, L\}$, where the dimensionality $N$ of the time series is much larger than the number $L$ of available observations. In its original formulation, DMD takes $x_t$ as a convenient state of an underlying process and thereby dispenses of higher order dynamics that may be hidden in differences between the time series data. The more general situation of higher order dynamics can be treated similarly \cite{le2017higher}. The main issue that we discuss in this paper is the pertinence of the assumption in seeking such a state model, and whether a reliable estimate of state dynamics should be expected to reflect the structure of the data. We propose a certain geometric concept, the so-called gap metric, as a tool to provide guidance in  selecting  suitable  dimension for the sought DMD dynamics.

\section{
The basic DMD rationale}

Consider the basic linear dynamical model,
\begin{equation}\label{eq:linear}
x_{t+1}=Ax_t + v_t, \mbox{ for } 1\leq t\leq L-1,
\end{equation}
where where $A\in\mathbb R^{N\times N}$, while $v_t\in\mathbb R^{N}$ signifies deviation from linear deterministic dynamics (via the input term $v_t$ that may represent stochastic excitation or contribution of nonlinear terms). The standard formulation of DMD is based on the assumption that the time series under consideration, herein $x_t$, is dominated by the linear transition mechanism and that, moreover, the dimension of $x_t$ is much larger than the size of the observation window $t\in\{1,2,\ldots, L\}$. 

The underlying premise of the DMD methodology is that the state vector $x_t$ concentrates along the directions that correspond to the dominant eigendirections of $A$ and, thereby, DMD aims (and has a viable chance) to identify the dynamics that are manifested by
restricting the recurrence relation in \eqref{eq:linear} onto the range of a data matrix
\[
X_{1:n-1}:=\left[ x_1,\,x_2,\,\ldots, x_{n-1}\right],
\]
for $n$ possibly $n\leq L$. Thereby, the dynamics are sought in a matrix
 $A\in \mathbb R^{N\times N}$ to satisfy
\begin{equation}\label{eq:approximate}
X_{2:n}\simeq AX_{1:n-1}.
\end{equation}

One readily observes that the operator $A$, restricted onto the orthogonal complement
of $\range (X_{1:n-1})$, namely,
\[
A|_{\range (X_{1:n-1})^\perp},
\]
is undefined, i.e., it cannot be determined from the data. DMD sets out to determine the action of $A$ precisely on the range of $X_{1:n-1}$.
To this end, complete the columns of $X_{1:n-1}$ into a basis $\mathcal B:=\{x_1,\ldots,x_{n-1},y_n,\ldots,y_N\}$ for $\mathbb R^N$. We tacitly assume that $\{x_1,\ldots,x_{n-1}\}$ are linearly independent. The matrix $Y_{n:N}=[y_n,\ldots,y_N]$ formed out of the added (column) vectors is such that
\[
T=\left[ X_{1:n-1}, Y_{n:N}\right]
\]
is an invertible matrix. Selection of $Y_{n:N}$ can be accomplished by taking the singular value decomposition
\[
X_{1:n-1} = U\Sigma V^T,
\]
of $X_{1:n-1}$, where $U\in O(N)$, $V\in O(n-1)$, and
\[
\Sigma=\left[\begin{matrix} \sigma_1(X_{1:n-1}) &0 &0 &\ldots\\
0& \sigma_2(X_{1:n-1}) & 0 & \ldots\\
0& 0& \ddots & \\
\vdots &\vdots &
 \end{matrix}\right]
\]
is the $N\times (n-1)$ matrix with the (non-increasing sequence of) singular values of $X_{1:n-1}$ on the main diagonal, and where $O(k)$ denotes the group of $k\times k$ orthogonal matrices. Then, if after partitioning
\[
U=\left[ U_{1:n-1}, U_{n:N}\right],
\]
the selection $Y_{n:N}=U_{n:N}$ presents a convenient option.

Similarity transformation with $T$ bring $A$ into the form
\[
\left[ \begin{matrix} S & S_{12}\\S_{21} &S_{22}\end{matrix}\right],
\]
since
\[
A \left[ X_{1:n-1}, Y_{n:N}\right] = \left[ X_{1:n-1}, Y_{n:N}\right]
\left[ \begin{matrix} S & S_{12}\\S_{21} &S_{22}\end{matrix}\right].
\]
Thus,
\[
AX_{1:n-1} = X_{1:n-1}S + Y_{n:N}S_{21}.
\]
Assuming that 
$A$ leaves
$\range (X_{1:n-1})$ invariant, the intertwining relation
\[
A X_{1:n-1} = X_{1:n-1} S,
\]
holds and $S$ represents the restriction of $A$ onto the $\range (X_{1:n-1})$. Thus, assuming that \eqref{eq:approximate} holds with equality,
\begin{equation}\label{eq:exact}
    X_{2:n} = X_{1:n-1} S,
\end{equation}
captures the action of $A$ on the range of $X_{1:n-1}$ and can be used to determine $S$.
Finally, because the columns of $X_{2:n-1}$ are shared with a shift between $X_{1:n-1}$ and $X_{2:n}$, $S$ has the companion structure
\[
S=\left[\begin{matrix} 0 & 0& 0 &\ldots &0 & -s_{n-1}\\
1& 0 &0 & \ldots & 0& -s_{n-2}\\
0 & 1 & 0 & \ldots &0 &-s_{n-3}\\
\vdots & &\ddots & && \vdots\\
\\
0& 0 & 0&\dots &1& -s_1\end{matrix}\right],
\]
where the last column can be easily identified by solving \eqref{eq:exact}.

Since in general the linear transformation $A$ does not leave $\range (X_{1:n-1})$ entirely invariant, and thereby \eqref{eq:approximate} does not hold with equality,
suitable approximation is carried out to obtain $S$. For instance, the vector $s=(s_{n-1},\ldots,s_1)^T$ can be
obtained as
\begin{equation}\label{eq:computation}
   {\rm argmin}\{ \|x_n-X_{1:n-1}s\|\mid s\in \mathbb R^{n-1}\},
\end{equation}
with $\|\cdot\|$ denoting (typically, and herein) the Euclidean norm, and to this end several alternative numerical schemes have proposed (such as Arnoldi and SVD based) \cite{schmid2010dynamic,kutz2016dynamic}. This is the typical scenario for DMD applications.

\subsection*{Regularizations}

An alternative approach is to regularize the problem by penalizing perturbation from the recorded values in data matrix $X_{1:n-1}$ as well, e.g., by solving instead (the nonlinear problem)
\[
{\rm argmin}\{ \|x_n-\hat X_{1:n-1} s\| + \epsilon\|\hat X_{1:n-1}- X_{1:n-1} \| \},
\]
over $s\in \mathbb R^{n-1}$ and $\hat X_{1:n-1}\in\mathbb R^{N\times (n-1)}$, for a choice of regularizing parameter $\epsilon>0$. This option is especially reasonable in case \eqref{eq:approximate} fails to hold with equality due to stochastic noise or the (small) effect of nonlinear dynamics, or in cases where prior information dictates specific structural features, e.g., see \cite{jovanovic2014sparsity,dicle2016robust}.

\subsection*{Higher order dynamics}
We note that in cases when higher order dynamics are at play and
$\mathbb R^N$ is insufficient as a choice of state-space, an option is
to account for lagged values of $x_t$ and thereby select as a candidate state vector, e.g., for the case of one lag,
\[
\xi_t = [x_t^T,\; x_{t-1}^T]^T.
\]
Very little changes in the basic setting \cite{le2017higher}. In this case, one seeks an $A$ matrix of twice the size to now satisfy $\Xi_{3:n}\simeq A\Xi_{2:n-1}$, cf.\ \eqref{eq:approximate}. Thence, a matrix $S$ as before, with companion structure, such that
\[
    \Xi_{3:n} = \Xi_{2:n-1} S,
\]
with $\Xi_{k:\ell}:=\left[ \xi_k,\,\xi_{k+1},\,\ldots, \xi_\ell\right]$, assuming $k<\ell$, cf.\ \eqref{eq:exact}. Thus, without loss of generality we will only discuss the basic setting without further expanding neither on higher order dynamics nor on the relevance of various choices for regularization.

\subsection*{Recap \& concluding thoughts}
The goal of DMD is to identify dominant modes that capture the relation between successive vectors of the time series. These are the roots
of the polynomial
\[
\mathbf s(\lambda)=\lambda^{n-1} +s_1 \lambda^{n-2}+\ldots +s_{n-1}.
\]
An underlying premise of the framework is that the time series does not depart significantly from being quasi-stationary.
This can only hold if the observed dynamics result in from a ``tug-of-war'' mechanism that provides excitation and saturation at the same time ({\em a la} fluctuation-dissipation). Such a dynamical mechanism can be based in either or both, a stochastic excitation or nonlinear contributions, as in \eqref{eq:linear}, where $v_t$ may represent either. This understanding suggests that the effectiveness of DMD and relevance of the underlying dynamical structure may be quantified by the geometric relation between subspaces spanned by successive collections of time series samples $x_t$. From a more practical perspective, the effectiveness of DMD, by necessity, rests on how close the subspaces spanned by $X_{2:n}$ and $X_{1:n-1}$ are.

To this end, below, we explore the use of geometric concepts that quantify how well the above expectations are reflected in the data. Specifically, we introduce the analogue of partial autocorrelation coefficients that can serve to identify the size of the state-space that can usefully be exploited to identify dominant dynamics.

\section{Innovation parameters (IP's)}

The effectiveness of DMD in modeling the underlying dynamics rests on the relation between the subspaces spanned by $\{x_\ell,x_{\ell+1},\ldots,x_m\}$, over a progression of intervals $[\ell,m]$ of indices and over varying window sizes.

Consider first intervals $[1,k-1]$ and $[2,k]$. We seek to quantify
the new information that is contained in the last vector $x_k$ as compared to the previous ones. Specifically, we consider how introducing these new data point $x_{k}$ impacts the distance (angle) between the subspaces spanned by $X_{1:k-1}$ and $X_{2:k}$.
Evidently, the angle between these subspaces relates to the discrepancy in \eqref{eq:approximate} from holding with identity.

We will similarly consider relations between subspaces corresponding to adjacent windows
$[\ell,\ell+k-1]$ and $[\ell+1,\ell+k]$, and how angles between such subspaces change with the indices $\ell$ and $k$.

\subsection*{ }
The distance between subspaces $\mathcal X_1, \mathcal X_2\subseteq \mathcal X$, of a Hilbert space $\mathcal X$, is naturally quantified by the angle operator
\[
R_{12}:= \Pi_{\mathcal X_1}|_{\mathcal X_2^\perp},
\]
where $\Pi_{\mathcal X_1}$ denotes orthogonal projection onto ${\mathcal X_1}$ and $|_{\mathcal X_1^\perp}$ the restriction onto the orthogonal complement of $\mathcal X_2^\perp$. Herein, we will be concerned with finite dimensional Euclidean spaces. In this case,
provided the subspaces have equal dimension,
\[
\| \Pi_{\mathcal X_1}|_{\mathcal X_2^\perp}\| =\| \Pi_{\mathcal X_2}|_{\mathcal X_1^\perp}\|.
\]
This common value is equal to $\|\Pi_{\mathcal X_1}-\Pi_{\mathcal X_2}\|$ and defines a {\em bona fide} metric between subspaces \cite{stewart1990matrix,kato2013perturbation}. This is referred to as the {\em
gap metric}
\[
d(\mathcal X_1, \mathcal X_2):=\|\Pi_{\mathcal X_1}-\Pi_{\mathcal X_2}\|.
\]
Thence, 
\[
\theta(\mathcal X_1, \mathcal X_2):= {\rm arcsin}(d(\mathcal X_1, \mathcal X_2))
\]
represents an angular distance between the two subspaces. In case their dimensions do not match, the gap is the maximal norm of the two angle operators, and equals $d(\mathcal X_1, \mathcal X_2)=1$, giving $\theta(\mathcal X_1, \mathcal X_2)=\frac{\pi}{2}$.

We remark that the gap metric between the graphs (infinite dimensional subspaces) of dynamical systems is a natural metric to quantify uncertainty in the context of feedback theory, and as such has been a chapter in modern robust control \cite{georgiou1988computation,georgiou1989optimal,zhou1998essentials}. Herein we are only concerned with the geometry of finite dimensional subspaces spanned by the vectorial entries of a time series.

\subsection*{Innovation parameters and PARCOR's}
In order to assess the consistency of successive measurements of the time series we consider gaps between subspaces spanned by successive segments, e.g., ${\rm range}(X_{1:k})$ and ${\rm range}(X_{2:k+1})$ for different values of $k$. We refer to these as
{\it innovation parameters (IP)}
\[
\mbox{\fbox{$r_k := d({\rm range}(X_{1:k}),{\rm range}(X_{2:k+1})$)}}
\]
In geometric terms, $r_k$ is the sine of the angular distance
\[
\theta_k:={\rm arcsin}(r_k)
\]
between $\Pi_{\thespan(x_2,\ldots,x_k)^\perp} x_1$ and $\Pi_{\thespan(x_2,\ldots,x_k)^\perp} x_{k+1}$, i.e., between the projections of $x_1,x_{k+1}$ onto the orthogonal complement of the span of the intermediate vectors $\{x_1,\ldots,x_{k}\}$. Similarly, we define
\[
\mbox{\fbox{$r_{\ell,k} := d({\rm range}(X_{\ell:\ell+k-1}),{\rm range}(X_{\ell+1:\ell+k})$)}}
\]
to capture the same dependence between successive subspaces from a different starting point $\ell$.

The innovation parameters relate to the partial correlation coefficients (PARCOR) in time-series analysis \cite{stoica2005spectral}. Specifically, if
\newcommand{\bX}{{\mathbf X}}
$\bX_k$, for $k\in\mathbb Z$, denotes a stationary time series, the PARCOR's
are the cosines of the angles between
\begin{align*}
& \bX_\ell-\mathbb E\{\bX_\ell|\bX_{\ell+1},\ldots,\bX_{\ell+k-1}\} \mbox{ and }\\
& \bX_{\ell+k}-\mathbb E\{\bX_{\ell+k}|\bX_{\ell+1},\ldots,\bX_{\ell+k-1}\},
\end{align*}
where in the conditioning, for $k=1$, we define the set $\{\bX_{\ell+1},\ldots,\bX_{\ell+k-1}\}$ as empty.
Thus, these also coincide with the cosines of the angles between the spans of the random variables $\{\bX_\ell,\ldots,\bX_{\ell+k-1}\}$ and $\{\bX_{\ell+1},\ldots,\bX_{\ell+k}\}$.

Besides one set of parameters corresponding to sines and the other to cosines, the main difference between IP's and PARCORs is that the latter are typically defined for stationary stochastic processes, in that the kernel
\[
\mathcal K(i,j):=\langle x_i,x_j\rangle
\]
in the case of IP's does not have necessarily a Toeplitz structure, as in the context where PARCORs \cite{stoica2005spectral}; the geometric relations in the data sequence $x_1,x_2,\ldots$ are not shift-invariant, in general, which often necessitates exploring the double indexing in $r_{\ell,k}$.

\subsection*{Recursive computation of innovation parameters}

Efficient code for computing the innovation parameters for large data sets and size of vectors can be devised based on a recursive scheme
that orthonormalizes successive vectors in the data base.

Specifically,
consider a basis for the span of $X_{1:k-1}$ to consist of $x_1$ and the orthonormal columns of a matrix $U_{2:k-1}$.
Likewise, the span of $X_{2:k}$ consist of $x_{k}$ and the orthonormal columns of a matrix $U_{2:k-1}$.
Define the orthogonal projection onto the orthogonal complement of the range of $U_{2:k-1}$
\[
\Pi_{{\rm range}(U_{2:k-1})^\perp}= I-U_{2:k-1}U_{2:k-1}^T.
\]
Then the angle between the span of $X_{1:k-1}$  and that of $X_{2:k}$ coincides with the angle between
\[
(I-\Pi_{{\rm range}(U_{2:k-1})^\perp})x_1 \mbox{ and }(I-\Pi_{{\rm range}(U_{2:k-1})^\perp})x_k.
\]
The computation of the innovation parameters can be carried our recursively as follows:\\
\begin{algorithm}
\SetAlgoLined
{{\bf Data:} Given $X_{1:n}\in\mathbb R^{N\times n}$}\\
{\bf Initialization:} $k=1$, $u_1=x_1/\|x_1\|$, $u_2=x_2/\|x_2\|$,\\
 $u_{\rm first}=u_1-\langle u_1,u_2\rangle u_2 $, $u_{\rm last}=u_2$, $U=[u_2]$\;
 \While{While $k<n-1$}{
 $u_{\rm last}=x_{k+2}$\;
 $u_{\rm last}=u_{\rm last}-UU'u_{\rm last}$\;
 $u_{\rm last}=u_{\rm last}/\|u_{\rm last}\|$\;
 $r_k=\sin({\rm acos}(\langle u_{\rm first},u_{\rm last}\rangle))$\;
 $U=\left[U ~~ u_{\rm last}\right]$\;
 $u_{\rm first}=u_{\rm first}-\langle u_{\rm first},u_{\rm last}\rangle u_{\rm last}$\;
 $u_{\rm first}=u_{\rm first}/\|u_{\rm first}\|$\;
  $k=k+1$\;
 }
 \caption{Recursive computation of IP's}
\end{algorithm}

Alternatively, the same computation can be carried out in Matlab utilizing the ``econ'' feature that optimizes computations for large data sets. E.g., in order to compute $r_n$ set $Y_1=X_{1:n-1}$ and $Y_2=X_{2:n}$, and compute $U_i$ for $i\in\{1,2\}$ with the command $ [U_i,\Sigma_i,V_i]={\rm svd}(Y_i,'{\rm econ}')$.
Since,
\[
\Pi_{\range(Y_i)}=U_iU_i',
\]
with $U_i$ an isometry, 
the gap between the two subspaces is
\begin{align*}
\| U_1 U_1' (I-U_2 U_2')\|^2&= \| U_1'-\underbrace{(U_1'U_2)}_{M} U_2'\|^2\\
& =\|  (U_1'-M U_2' )( U_1-U_2 M')\| \\
& = \| I - MM'- MM' + MM'\|\\
&= \|I-MM'\|
\end{align*}
Therefore, the gap between $\range(Y_1)$ and $\range(Y_2)$ is
\[
\sqrt{1 - \sigma_{\min}(M)^2}
\]
with $M=U_1^\prime U_2$.

We proceed to motivate and explain the use and relevance of the IP's in selecting a suitable size $n$ for the dynamics sought via DMD on a case study. An additional technical result will be presented along with the example, which highlights the fact that under- or over-estimating the value for $n$ leads to significant errors in identifying the correct dynamics. The example we consider is that of an almost periodic series.

\section{A case study}

\begin{figure}[htbp]
\centering
\subfigure[$t=1$]{\centering
\includegraphics[width=7cm]{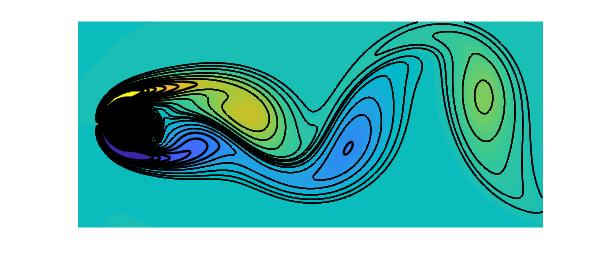}}
\hfill
\subfigure[$t=5$]{\centering
\includegraphics[width=7cm]{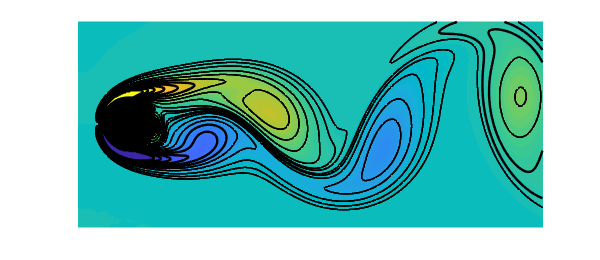}}
\hfill
\subfigure[$t=9$]{\centering
\includegraphics[width=7cm]{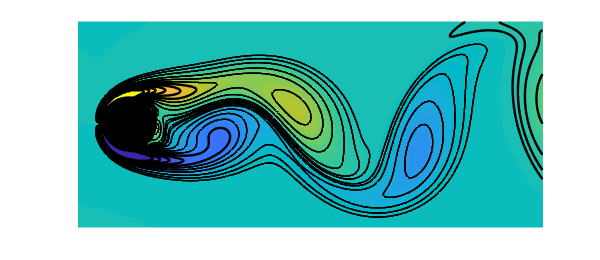}}
\caption{Vorticity field around a cylinder wake.}
\label{Cylinder_wake}
\end{figure}

We consider time series data that represent a persistent vorticity of a periodically fluctuating fluid flow
field in the wake behind a circular cylinder. This dataset can be generated by publicly accessible code in \cite{cwrowley}. The two-dimensional Navier–Stokes equations are numerically solved
at Reynolds number 100, to obtain these data. At this Reynolds number the flow undergoes a laminar vortex shedding which can be thought of as a stable limit cycle. The data are collected after simulations converge to steady-state vortex shedding. The reader is referred to \cite{kutz2016dynamic} for more details on how these data set is extracted. At each of $151$ snapshots, the values of vorticity are stacked up in a column of a data matrix $X$ which is of size $89351\times 151$. The images of the vorticity field at successive timestamps $t\in\{1,5,9\}$ are depicted in Fig. \ref{Cylinder_wake}. The color-coded velocity fluctuations reveal the mechanism of vortex shedding.

The DMD formalism, and specifically \eqref{eq:computation}, is applied to identify the apparent modes of oscillation. The resulting modes are dramatically affected by the choice of $n$ in \eqref{eq:computation}.
Important points that are highlighted below by this example are as follows:
\begin{itemize}
\item[i)] The time series is very close to being periodic. This can been seen in a variety of ways, including standard spectral or Fourier analysis. However, here, we compute the sequence of innovation parameters that quantify how far the subspaces spanned by sliding windows of data, of varying width, are from each other in the gap metric.

\begin{figure}[htbp]
\begin{center}
\includegraphics[width=7.5cm]{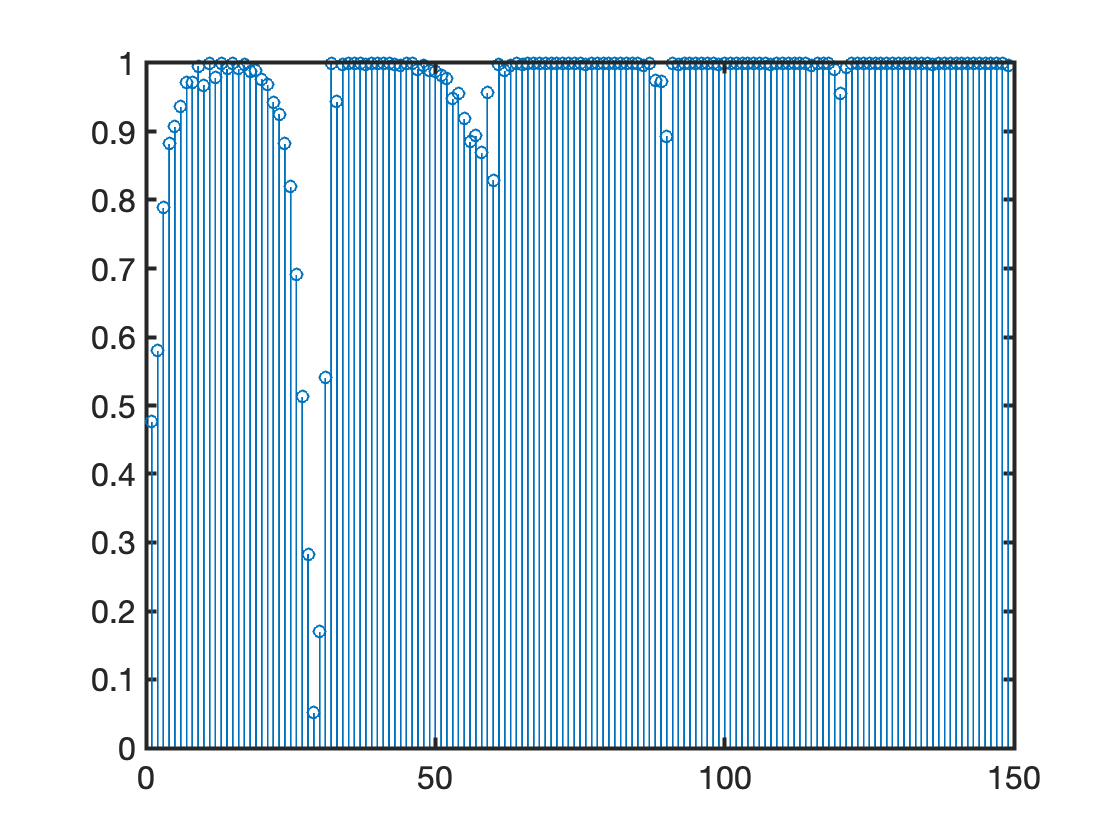}
\caption{$r_k=d({\rm range}(X_{1:k}),{\rm range}(X_{2:k+1}))$ vs.\ $k$}
\label{ip_wake}
\end{center}
\end{figure}

\begin{figure}[htbp]
\begin{center}
\includegraphics[width=7.5cm]{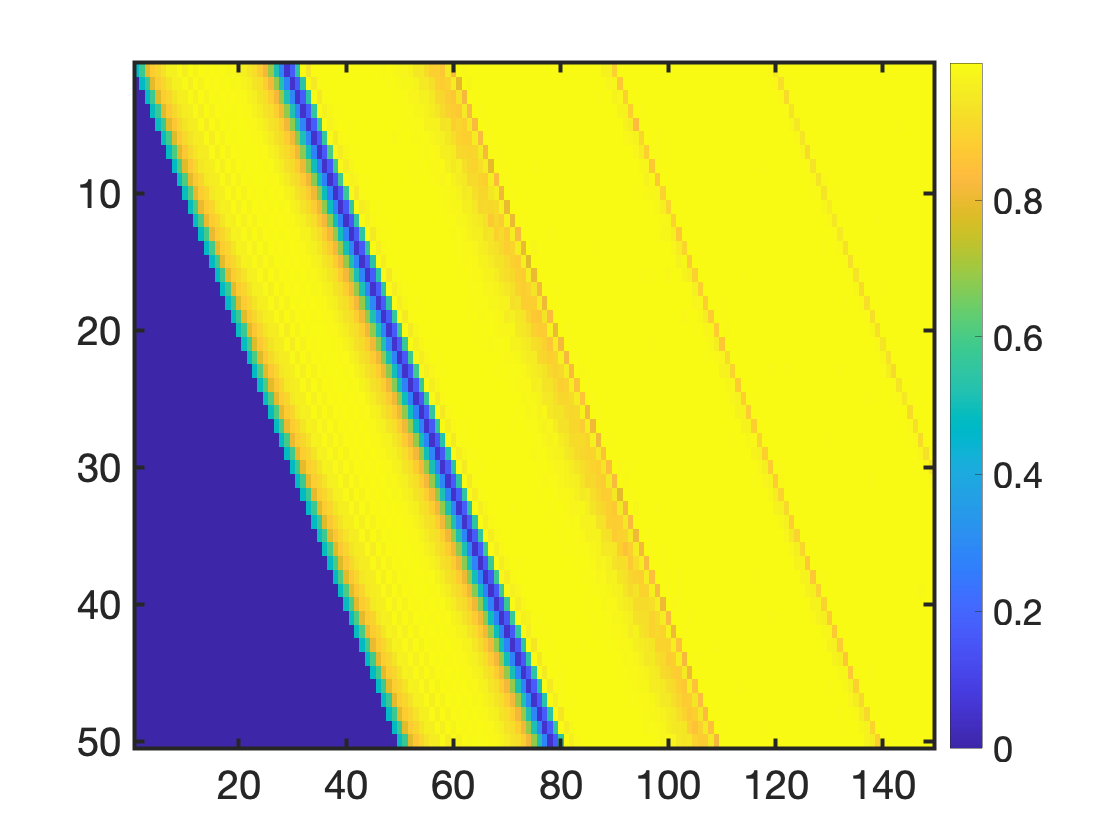}
\caption{$d({\rm range}(X_{\ell:k}),{\rm range}(X_{\ell:k+1}))$ color-coded as function of starting time $\ell$ and window size $k$}
\label{spectogram}
\end{center}
\end{figure}

Fig.\ \ref{ip_wake} shows $r_k=r_{1,k}$ as a function of $k$. A dimple that repeats with period $30$ indicates periodicity. It turns out that exact periodicity of the $r_k$'s, even when the time series is very close to being periodic is masked by numerical sensitivity that we will comment later on (discussion leading to, and Proposition \ref{prop:1}).

\item[ii)] Fig.\ \ref{spectogram} shows the color-coded values of $r_{\ell,k}$ as a function of $\ell$ vs.\ $k$.  Specificaly, $50$ snapshots are drawn as rows. The $i\text{th}$ row corresponds to  the gap $d({\rm range}(X_{\ell:k}),{\rm range}(X_{\ell:k+1}))$, where $k$ sweeps from $\ell+1$ to the last one. The first row, for instance, corresponds to the values illustrated in Fig.\ \ref{ip_wake}.
One can observe that at each row the minimum gap occurs at the 30th timestamp. This strongly suggests the use of a time-window of size $n=30$ to find the DMD modes. 
Periodicity is evident in Fig.\ \ref{spectogram}; the decreasing dimples with period $30$ are repeated with regularity starting from any chosen starting point $\ell$ (cf.\ discussion leading to Proposition \ref{prop:1}).

\begin{figure}[htbp]
\centering
\subfigure[DMD eigenvalues for $n=20$]{\centering
\includegraphics[width=7cm]{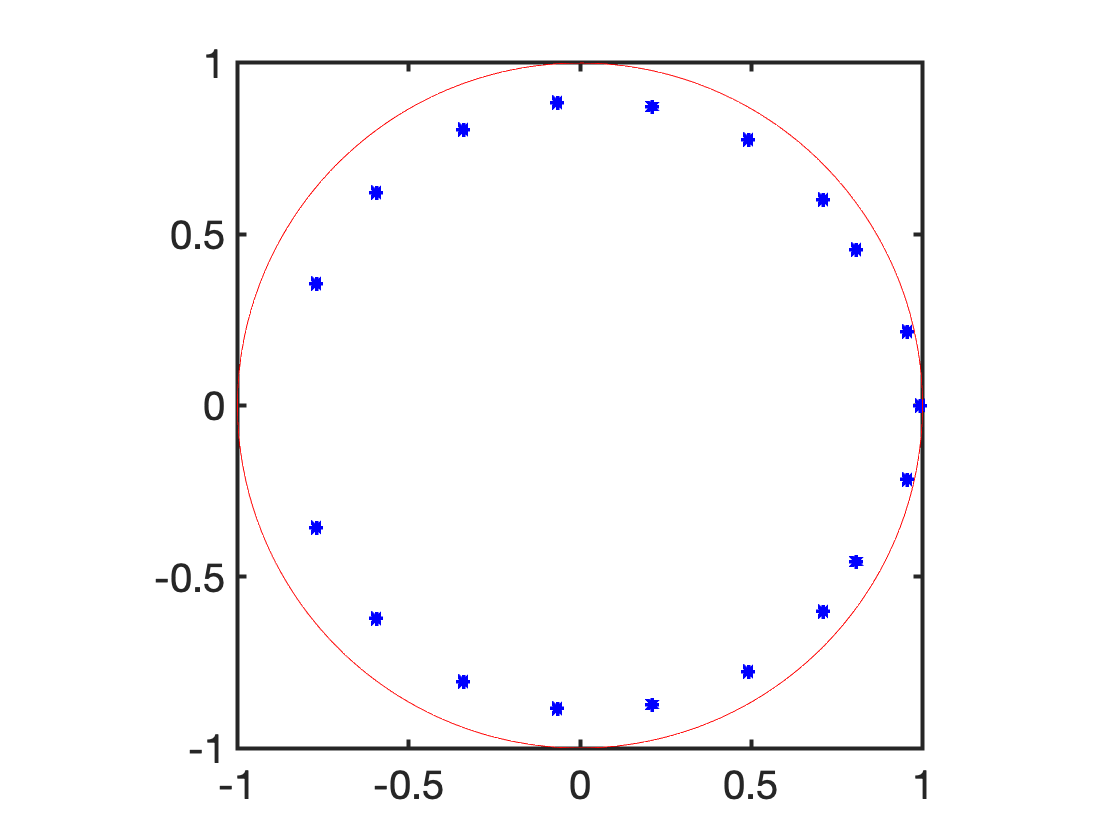}}
\subfigure[DMD eigenvalues for $n=30$]{\centering
\includegraphics[width=7cm]{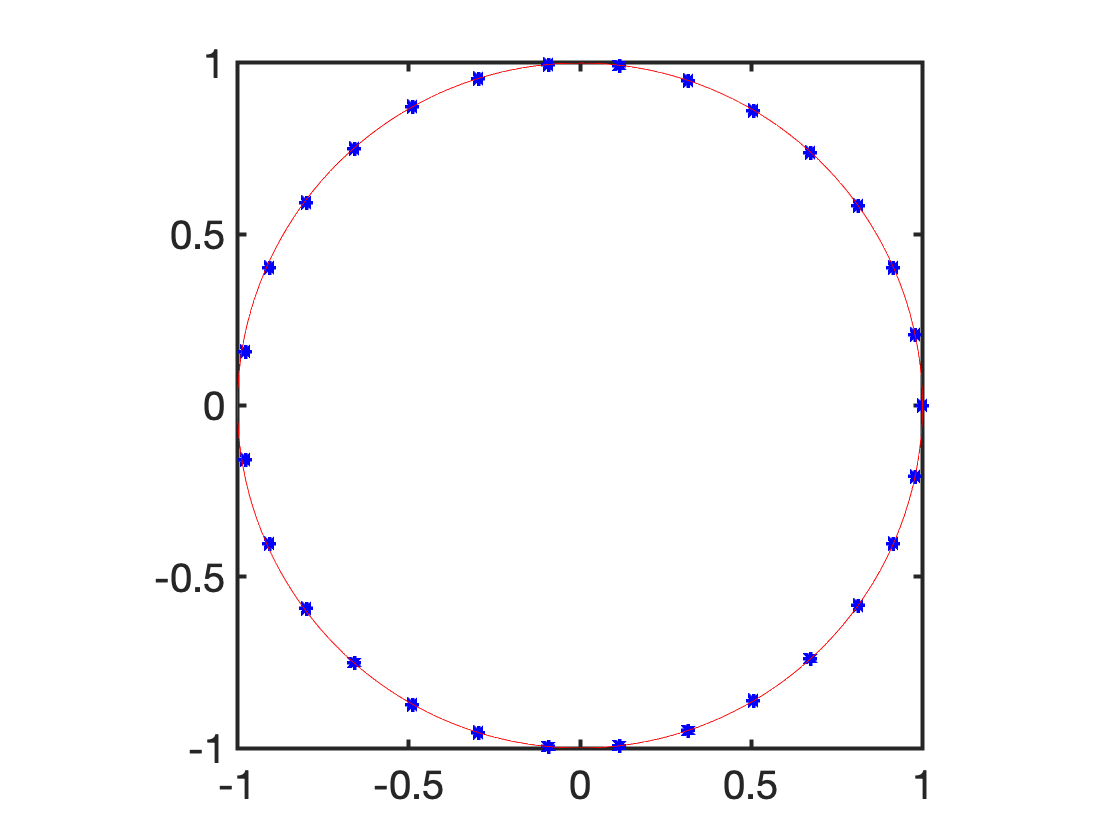}}
\subfigure[DMD eigenvalues for $n=40$]{\centering
\includegraphics[width=7cm]{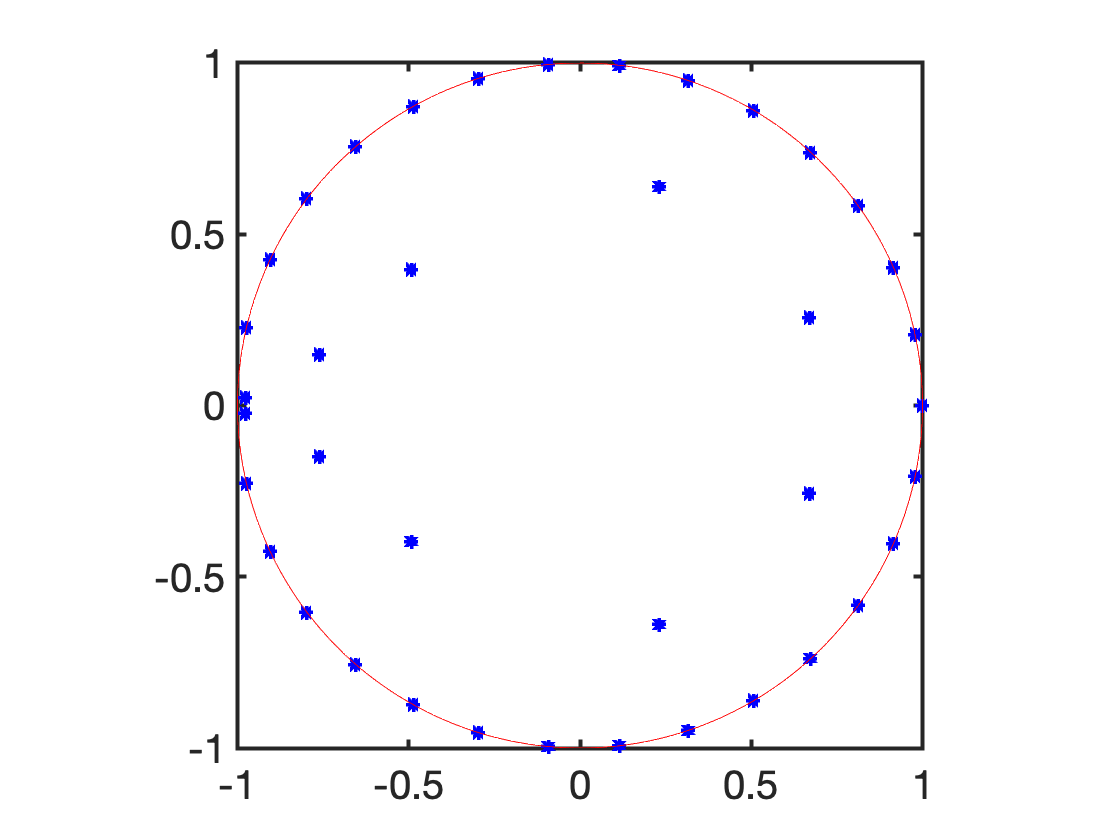}}
\caption{DMD eigenvalues from vorticity field data.}
\label{Cylinder_eigenvalues}
\end{figure}

\item[iii)] The eigenvalues of $S$ (DMD eigenvalues of the sought dynamics) are shown in Fig.\ \ref{Cylinder_eigenvalues}
for $n\in\{20,\,30,\,40 \}$. It is observed that their distribution is dramatically affected by the chosen size of the subspaces to compare in \eqref{eq:approximate}, namely $n$.

\item[iv)] For $n=30$ the eigenvalues of $S$ shown in Fig.\ \ref{Cylinder_eigenvalues} have modulus $\simeq 1$, in agreement with the observed periodic structure of the flow field. Exact periodicity of the time series results in equispaced eigenvalues, and this is (almost) the case here.

\end{itemize}

At this point we would like to explain the source of the apparent diminishing of periodic dimples in Fig.\ \ref{ip_wake} with period $30$. As noted earlier, the gap 
\[
r_k=d({\rm range}(X_{1:k-1}),{\rm range}(X_{2:k}))
\]
is the sine of the angle between 
\begin{align*}
\xi_1&:=\Pi_{\thespan(x_2,\ldots,x_{k-1})^\perp} x_1, \mbox{ and}\\
\xi_{k}&:=\Pi_{\thespan(x_2,\ldots,x_{k-1})^\perp} x_{k}.
\end{align*}
Assuming that the series is $k$-periodic, the angle between $\xi_1$ and $\xi_k$ is  zero and $x_k\in \thespan(x_1,\ldots, x_{k-1})$. Likewise, 
\[
x_{k+1}\in \thespan(x_2,\ldots, x_{k})=\thespan(x_1,\ldots, x_{k-1}).
\]
Denote
\[
\xi_{\rm next}:=\Pi_{\thespan(x_2,\ldots,x_{k-1})^\perp} x_{k+1},
\]
and observe that the angle to $\xi_k$, and therefore $\xi_1$ too, is zero.
Then
\begin{align*}
r_{k+1}&=d({\rm range}(X_{1:k}),{\rm range}(X_{2:k+1}))\\
&=d({\rm span}(\xi_1,X_{2:k-1},\xi_k),{\rm span}(X_{2:k-1},\xi_k,\xi_{\rm next}))\\
&=d({\rm span}(\xi_1,\xi_k),{\rm span}(\xi_k,\xi_{\rm next}))=0,
\end{align*}
with all three vectors $\xi_1,\xi_k,\xi_{\rm next}$ co-linear.
However, a small perturbation in each has a significant effect. Indeed, for arbitrarily small $\delta$'s,
\begin{align*}
&d({\rm span}(\xi_1+\delta_1,\xi_k+\delta_k),{\rm span}(\xi_k+\delta_k,\xi_{\rm next}+\delta))\\
&=d(\Pi_{\thespan(\xi_k+\delta_k)^\perp} (\xi_1+\delta_1),\Pi_{\thespan(\xi_k+\delta)^\perp} (\xi_{\rm next}+\delta))
\end{align*}
can take any value on $[0,1]$.
We recast the claim as follows.
\begin{prop}\label{prop:1}
Consider a vector $\xi\in\mathbb R^N$ and perturbations $\xi_i=\xi+\delta_i$, for $i\in\{1,2\}$, with $\delta_i \perp \xi$.
Then
\[
d({\rm span}(\xi+\delta_1,\xi),{\rm span}(\xi+\delta_2,\xi)=d({\rm span}(\delta_1),{\rm span}(\delta_2)).
\]
\end{prop}
The proof is elementary. What this statement helps exemplify (and prove) is that in  cases where elements that determine the span of interest are almost co-linear, the angles between the subspaces are very sensitive to errors. A more precise mathematical statement can be worked out that involves the conditioning number of the matrix $X_{1:k}$ in our earlier setting.

\section{Concluding remarks}

In many applications it is often the case that only a limited number of data samples are available for modeling an otherwise exceedingly high dimensional process. The dimensionality of the process, which may represent visual or distributional fields, in conjunction with the limited observation record requires careful analysis. It is precisely this regime of ``small data,'' i.e., ``few samples,'' that has been a challenge in traditional signal analysis since its inception \cite{burg1982estimation}, and has led to entropic regularization among other methodologies. DMD represents a more recent development that aims to identify suitable linear dynamics that can explain the data.

Historically, DMD has roots and ramifications that relate to theory of the Koopman operator \cite{mezic2000comparison,mezic2005spectral,rowley2009spectral}. Data that originate from periodic and quasi-periodic attractors of nonlinear dynamics can also be dealt with in the same framework \cite{kutz2016dynamic}.
Thus the concept of the gap metric, as a tool to quantify how subspaces spanned by data impact modeling assumptions, is expected to be applicable in this more general setting.
The present work summarizes some of the findings in a developing treatise into the topic of extracting dynamics from high dimension distributional fields \cite{AmirK}, specifically, the relevance of the gap metric as a tool to provide guidance in selecting appropriate dimensionality for models for such processes.

\bibliographystyle{plain}        
\bibliography{refs}      
\end{document}